\documentclass[twocolumn,superscriptaddress,amsmath,amssymb,pra]{revtex4-2}
\usepackage[utf8]{inputenc}
\usepackage[T1]{fontenc}
\usepackage{graphicx}
\usepackage{dcolumn}
\usepackage{appendix}
\usepackage{bm}
\usepackage{ulem}
\usepackage[usenames,dvipsnames]{xcolor}

\usepackage{subfigure}
\usepackage{bbm}
\usepackage{enumerate}

\usepackage[
  bookmarks=true,
  colorlinks,
  linkcolor=black,
  urlcolor=black,
  citecolor=black,
  plainpages=false,
  pdfpagelabels,
  final,
  breaklinks=true
]{hyperref}

\usepackage{titlesec}

\usepackage{array}

\usepackage{physics}

\begin{document}
\title{Nonlinear optics using intense optical coherent state superpositions}

\author{Th. Lamprou}
 \affiliation{Foundation for Research and Technology-Hellas, Institute of Electronic Structure \& Laser, GR-7001 Heraklion (Crete), Greece}

\author{J.~Rivera-Dean}
\affiliation{ICFO -- Institut de Ciencies Fotoniques, The Barcelona Institute of Science and Technology, 08860 Castelldefels (Barcelona)}

\author{P. Stammer}
\affiliation{ICFO -- Institut de Ciencies Fotoniques, The Barcelona Institute of Science and Technology, 08860 Castelldefels (Barcelona)}
\affiliation{Atominstitut, Technische Universit\"{a}t Wien, 1020 Vienna, Austria}

\author{M. Lewenstein}
\affiliation{ICFO -- Institut de Ciencies Fotoniques, The Barcelona Institute of Science and Technology, 08860 Castelldefels (Barcelona)}
\affiliation{ICREA, Pg. Llu\'{\i}s Companys 23, 08010 Barcelona, Spain}

\author{P. Tzallas}
\email{ptzallas@iesl.forth.gr}
\affiliation{Foundation for Research and Technology-Hellas, Institute of Electronic Structure \& Laser, GR-7001 Heraklion (Crete), Greece}
\affiliation{ELI-ALPS, ELI-Hu Non-Profit Ltd., Dugonics tér 13, H-6720 Szeged, Hungary}

\date{\today}

	\begin{abstract}
		Superpositions of coherent light states, are vital for quantum technologies. However, restrictions in existing state preparation and characterization schemes, in combination with decoherence effects, prevent their intensity enhancement and implementation in nonlinear optics. Here, by developing a decoherence--free approach, we generate intense femtosecond--duration infrared coherent state superpositions (CSS) with a mean photon number orders of magnitude higher than the existing CSS sources. We utilize them in nonlinear optics to drive the second harmonic generation process in an optical crystal. We experimentally and theoretically show that the non--classical nature of the intense infrared CSS is imprinted in the second-order autocorrelation traces. Additionally, theoretical analysis shows that the quantum features of the infrared CSS are also present in the generated second harmonic. The findings introduce the optical CSS into the realm of nonlinear quantum optics, opening up new paths in quantum information science and quantum light engineering by creating non-classical light states in various spectral regions via non-linear up-conversion processes. 
  
	\end{abstract}

\maketitle

Superpositions of coherent light states, correspond to an optical analog of the Schr\"{o}dinger's cat in his {\it Gedankenexperiment}~\cite{schrodinger_gegenwartige_1935}. A superposition composed by two coherent states of the same frequency, equal amplitude $\abs{\alpha}$ and opposite phase $\ket{\text{CSS}}_{\pm} = N_{\pm}(\ket{\alpha} \pm \ket{-\alpha})$ (where $N_{\pm}$ are the normalization factors) defines an optical CSS. Depending on the amplitude $|\alpha|$, we define the CSS
as "small", "medium" and "large" for $|\alpha| \lesssim 0.5$, $|\alpha| \approx 1$ and $|\alpha| \geq 2$, respectively, with the former being more robust against decoherence/losses~\cite{zhang_quantifying_2021} compared to the ``large'' CSS . In more general terms, we refer to superpositions of the form $\ket{\text{GCSS}} = A(\ket{\alpha_1} + \xi \ket{\alpha_2})$ ($A$, $\xi$ are the normalization factor and complex coefficient and $\ket{\alpha_{1,2}}$ are coherent states of the same frequency but different amplitudes), as generalized CSS (GCSS).  Coherent state superpositions have a vital role for the development of new quantum technologies. They are a powerful tool for fundamental tests of quantum theory\cite{sanders_entangled_1992,wenger_maximal_2003,garcia-patron_proposal_2004,stobinska_violation_2007}, and an unique resource for numerous of investigations in quantum information science \cite{giovannetti_quantum-enhanced_2004,gilchrist_schrodinger_2004,giovannetti_advances_2011,joo_quantum_2011, van_enk_entangled_2001,jeong_quantum-information_2001, lloyd_quantum_1999,ralph_quantum_2003,jouguet_experimental_2013, Cotte_PRR_2022}. Their generation largely relies on the implementation of optical methods combined with conditioning measurements \cite{zavatta_quantum--classical_2004,ourjoumtsev_generating_2006,ourjoumtsev_generation_2007,sychev_enlargement_2017,zhang_quantifying_2021, lewenstein_generation_2021,rivera-dean_strong_2022,stammer_high_2022,stammer_theory_2022,stammer_quantum_2023, Bhattacharya_2023, Gonoskov_PRB_2024, Javier_PRB_2024}. These notable techniques have reported the generation and characterization of GCSS with mean photon number in the range of few photons. However, despite their fundamental interest and utility, the low photon number of these states limits their applicability in many investigations within quantum technology and prevents their use in nonlinear optics.

The generation, characterization, and implementation of intense GCSS in nonlinear optics requires the development of: a) novel schemes capable of supporting the production of high power GCSS, b) methods that can support the detection of high photon number GCSS while minimizing decoherence effects induced by passive optical elements \cite{zhang_quantifying_2021, Teh_2020}, and c) new quantum light characterization methods. The latter is because the current characterization methods, such as quantum tomography or methods based on photon correlations \cite{Yuen_1983, Schumaker_1984, Boitier_2009, Boitier_2011}, face challenges in characterizing high mean photon number GCSS.  

Here, we demonstrate a scheme that overcomes these challenges and allows the generation, characterization, and implementation of intense GCSS in nonlinear optics. We generate and characterize GCSS in the infrared (IR) spectral region with mean photon number orders of magnitude higher than those offered by existing GCSS sources. We show that these states have practical applications in nonlinear optics and quantum light state engineering. In particular, we theoretically show that through their nonlinear interaction with an optical crystal, they lead to the generation of a second harmonic which depicts features of an GCSS. 

The scheme used to demonstrate the above is shown in Fig.~1 (Supplementary Matterial (SM) \cite{SM} Pts.1--7). A linearly polarized $\approx 25$ fs IR laser pulse (central wavelenght $\approx 800$ nm and bandwidth $\approx 40$ nm) with intensity $\sim 10^{14}$ W per cm$^{2}$ was focused into Argon (Ar) atoms. The interaction leads to high harmonic generation (HHG)~\cite{lewenstein_theory_1994,amini_symphony_2019}, where the low-frequency photons of a driving laser field are converted into photons of higher frequencies $q\omega$ ($q$ is the harmonic order). Odd harmonic orders with $q\leq35$ have been observed (SM \cite{SM} Pt.2).  According to the fully quantized theory of HHG ~\cite{lewenstein_generation_2021,rivera-dean_strong_2022,stammer_high_2022,stammer_theory_2022, stammer_quantum_2023} (SM \cite{SM} Pt.1), the IR and harmonic field states before the interaction are coherent, i.e.,  $\ket{\phi_{\text{I}}(t)}=\ket{\alpha_L(t)}\bigotimes_{q=2} \ket{0_q}$, with $\ket{\alpha_L(t)}=\ket{\alpha_L f(t)e^{i \omega t}}$. $f(t)$, $\omega$, $\abs{\alpha_L}$ are the envelope, the central frequency and the amplitude of the coherent state of the IR pulse. $\ket{0_q}$ represents the vacuum state of the $q$th ($q\geq 2$) harmonic mode. The interaction with Ar atoms leads to an amplitude shift in all field modes including the fundamental. This is described by the multi-mode displacement operator $\hat{D}(\delta\alpha_{L})\prod_{q}\hat{D}(\chi_{q})$. The shift of the fundamental and harmonic modes account for amplitude depletion $\delta \alpha_{L}$ ($\delta \alpha_{L}$ is negative due to IR energy losses) and harmonic generation $\chi_{q}$, respectively. Therefore, after the interaction the field state is in a product state $\ket{\phi(t)}=\ket{(\alpha_{L} + \delta \alpha_{L}) f(t) e^{i \omega t}} \bigotimes_{q}\ket{\chi_{q}}$, where the field modes are correlated \cite{Stammer_arXiv_2024, Stammer_arXivEnergy_2024}, as they are consequence of the same electron dynamics, i.e., the induced dipole moment which leads to HHG. We refer to the set of correlated field modes, including the fundamental and the harmonic modes, as $\{\ket{\tilde{n}}\}$, with $\Tilde{n}=0$ representing the absence of HHG excitations. 

\begin{figure*}
\centering
    \includegraphics[width=0.8 \textwidth]{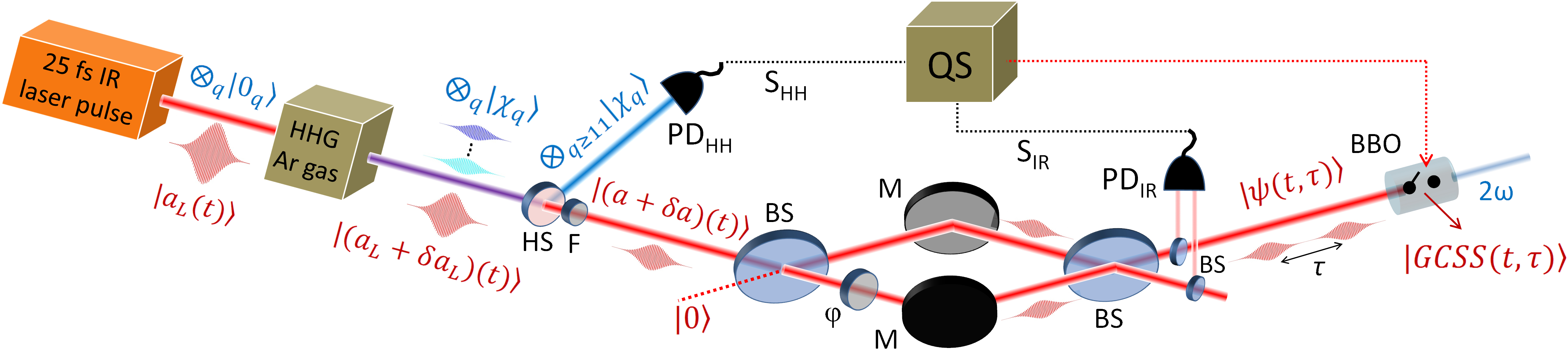}
    \caption{Scheme for utilizing intense optical GCSS in nonlinear optics. $\ket{\alpha_L(t)}\bigotimes_{q}\ket{0_q}$ is the initial coherent field state. $\ket{\alpha_L(t)}$ and $\ket{0_q}$ represent the IR laser and the $q$th harmonic vacuum states, respectively. The laser--Argon (Ar) interaction leads to high harmonic generation (HHG). $\ket{(\alpha_{L} + \delta \alpha_{L}) (t)} \bigotimes_{q}\ket{\chi_{q}}$ is the coherent state after the HHG. $\bigotimes_{q\geq11}\ket{\chi_{q}}$ are the harmonic states reaching the photodetector PD$_{\text{HH}}$. HS, F are harmonic separator, IR attenuator . The state $\ket{(\alpha + \delta \alpha) (t)}$ enters the Mach-Zehnder interferometer. M, BS are mirrors, beam splitters. $\varphi$ introduces the time delay $\tau$. The state $\ket{\psi (t, \tau)}$ generates the second harmonic ($2\omega$) in the BBO crystal. The QS creates, the optical GCSS $\ket{\text{GCSS}(t,\tau)}$ by projecting ($\mathbbm{1}-\dyad{\Tilde{0}}$) the $\ket{\psi_{(\text{IR},q)}}$ state on HHG. S$_{\text{HH}}$ and S$_{\text{IR}}$ are the photocurrents recorded by the PD$_{\text{HH}}$ and PD$_{\text{IR}}$ detectors and used by the QS for implementing the conditioning in a post selection process i.e., with the QS we post select on HHG to exclude unwanted nonlinear processes. We note that the physical process in the BBO crystal serves as a measurement tool for the IR GCSS state.}
        \label{fig:setup_scheme}
\end{figure*}

Then, the IR field is amplitude attenuated by the filter $F$, i.e. $\alpha_L \xrightarrow{} \alpha$ and $\delta\alpha_L \xrightarrow{} \delta\alpha$, resulting into $\ket{(\alpha + \delta \alpha) f(t) e^{i \omega t}}$, and subsequently directed through a Mach-Zehnder interferometer which introduces a phase shift $\varphi$ in one of the arms. This leads to a time-delay $\tau$ between the two coherent states which, inside the interferometer, are given by $\ket{\pm \frac{1}{\sqrt{2}}(\alpha + \delta \alpha)f_{\pm} e^{i \omega(t\pm \tau/2)}}$, where $f_{\pm}=f(t\pm\tau/2)$. These states overlap at the final beam-splitter (BS), such that the state propagating towards the BBO crystal is given by
\begin{equation} \label{eq:InterfState}
\begin{aligned}
\ket{\psi (t, \tau)}
    = \ket{ \frac{1}{2} (\alpha + \delta \alpha)
        \big[
            f_{+} e^{i \omega (t+ \tau/2)}
            + f_{-} e^{i \omega (t-\tau/2)}
        \big] }. 
\end{aligned}
\end{equation}
Incorporating the harmonic modes the total field state reads $\ket{\psi_{(\text{IR},q)} (t, \tau)}=\ket{\psi(t, \tau)} \bigotimes_{q}\ket{\chi_{q}}$.  The field described by Eq.~\eqref{eq:InterfState} is focused into the BBO crystal generating the second harmonic with frequency $2\omega$. The mean photon number of the $\ket{\psi (t, \tau)}$ state before the crystal is $\expval{n}=|\alpha + \delta \alpha|^2/4 \approx 1.5 \cdot 10^2$ photons per pulse, and the photon number of the $2\omega$ in the range of a few photons (SM \cite{SM} Pt.4). The $2\omega$ signal $S_{2\omega} (\tau) \propto \int\langle\hat{I}^2\rangle(t,\tau)dt$ is recorded as a function of the delay $\tau$ between the two fields. For the time-delayed coherent states of Eq. \eqref{eq:InterfState}, $S_{2\omega}(\tau)$ depicts the well known features of a conventional 2-AC trace produced by a pulsed coherent light state (Fig.~2a). 

\begin{figure}
\centering
    \includegraphics[width=1 \columnwidth]{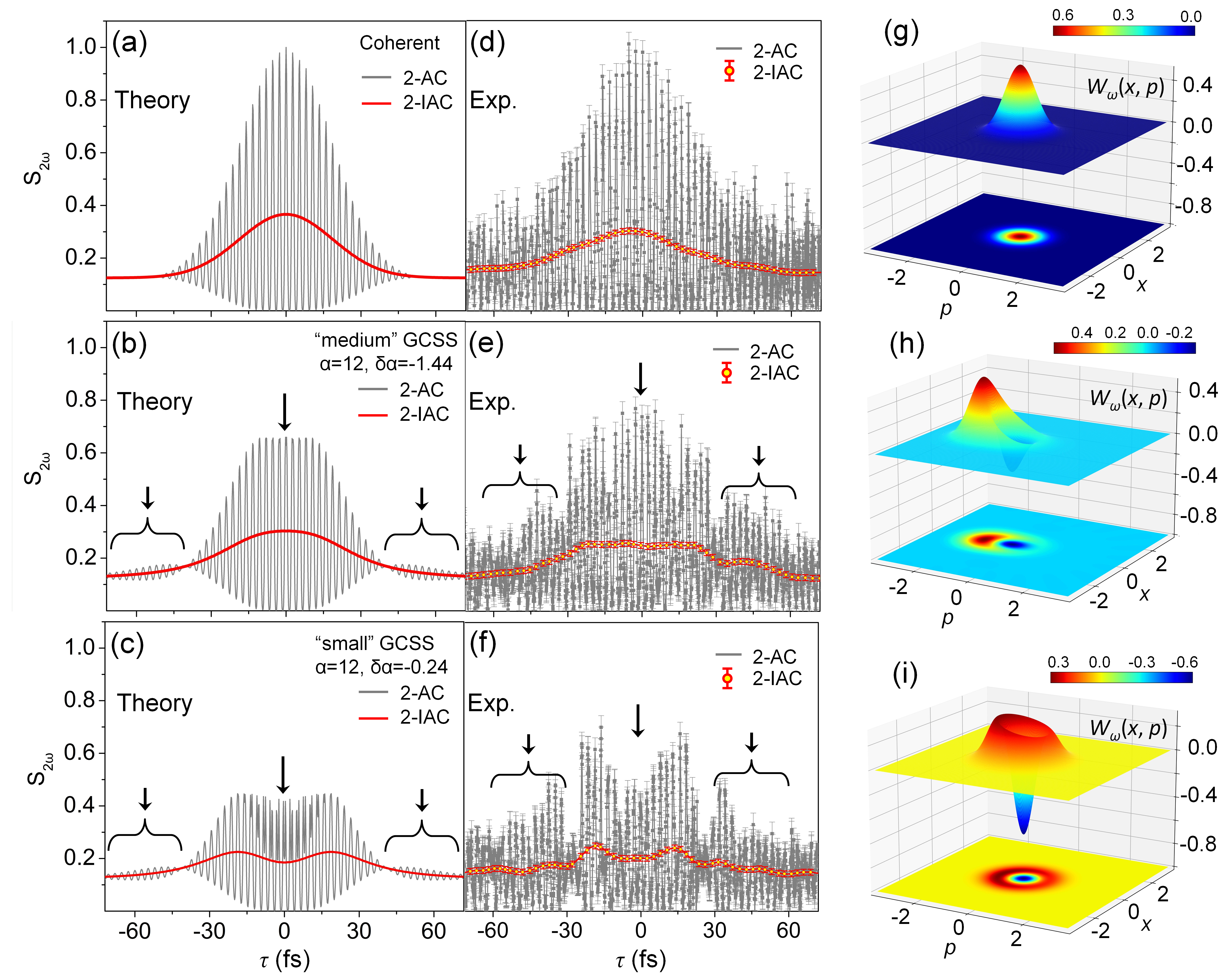}
    \caption{ Left panels ((a)--(c)), calculated second order interferometric autocorrelation (2--AC) traces. (a) When the QS is off the 2-AC trace corresponds to a conventional 2--AC trace of a coherent light pulse. (b), (c) 2-AC traces calculated when the QS is on and an ``medium'' and ``small'' GCSS with $|\delta\alpha|=1.44$ and $0.24$, respectively, is created. Middle panels ((d)--(f)), the corresponding to (a)--(c) experimentally measured 2--AC traces. In all traces the red lines (red yellow--filled circles) result from the cycle average of the 2-AC traces and they correspond to the second order intensity autocorrelation (2-IAC) traces. The error bar represents one standard deviation of the mean. The black--arrows depict the time delay regions where the quantum interference between the coherent states composing the GCSS is prominent. Right panels ((g)--(i)) Theoretical Wigner functions (centered around the value of $\alpha$) $W_{\omega}(x, p)$ in phase space $(x, p)$ of a coherent (g), ``medium'' GCSS with $|\delta\alpha|=1.44$ (h), and ``small'' GCSS with $|\delta\alpha|=0.24$ (i).}
        \label{fig:theory:main}
	\end{figure}

The situation drastically changes when we project the $\ket{\psi_{(\text{IR},q)}}$ onto the part that has been affected by the HHG, i.e. postselect (conditioning) on events where the depletion $\delta\alpha$ of the IR state $\ket{\psi (t, \tau)}$ leads to a shift $\chi_{q}$ of the harmonic modes represented by $\{\ket{\chi_{q}}\}_{q\geq2}$. In this case, as is expected from energy conservation, reduction of the IR amplitude (increase of the depletion $|\delta\alpha|$) leads to enhancement of the harmonic amplitudes $|\chi_{q}|$.  In this postselection procedure, for each shot $i$ we select the IR photons that have been absorbed from the driving field $\Delta n'_{\text{IR}}(i)=n_{\text{IR}}-n'_{\text{IR}}(i)$ towards the generation of $n_{\text{HH}}(i)$ harmonic photons, where $n_{\text{IR}}$ and $n'_{\text{IR}}(i)$ are the photon numbers of the undepleted and depleted IR fields respectively, and obey $n'_{\text{IR}}(i)\leq n_{\text{IR}}$. The exact expression of the conditioning operator is given in Ref. \cite{Stammer_arXiv_2024, Stammer_arXivEnergy_2024}, taking into account the shot to shot photon number measurement or the IR and harmonic photons. It has been found in Ref. \cite{Stammer_arXiv_2024, Stammer_arXivEnergy_2024} that the conditioning procedure is sufficiently described by a set of positive operator-valued measures (POVM) introduced in Refs.~\cite{stammer_high_2022, stammer_theory_2022}. This POVM consists of $\{\mathbbm{1}-\dyad{\Tilde{0}},\dyad{\Tilde{0}}\}$, describing the case when the HHG process and the corresponding IR depletion have occurred or not, respectively (SM \cite{SM} Pt.1). $\ket{\Tilde{0}} \equiv \ket{\alpha(t)} \bigotimes_{q} \ket{0_q}$ is the state without HHG excitations, i.e. the initial state before the interaction. We refer to $\hat{P}_{\text{HHG}} \equiv \mathbbm{1}-\dyad{\Tilde{0}}$ as the conditioning on HHG, leading to the state,
\begin{equation} \label{eq:PhiState}
\begin{aligned}
\ket{\Phi_{(\text{IR},q)} (t,\tau)}         
    &=\hat{P}_{\text{HHG}}\ket{\psi_{(\text{IR},q)} (t, \tau)}\\
    &=\ket{\psi (t, \tau)}\bigotimes_{q} \ket{\chi_q}- \Tilde{\xi}
    (t,\tau)\ket{\alpha (t)}\bigotimes_{q} \ket{0_q},
\end{aligned}
\end{equation}
where $\Tilde{\xi}(t,\tau) \equiv \xi_{\text{IR}}(t,\tau) \xi_{q}$, with $\xi_{\text{IR}}(t,\tau)=\bra{\alpha (t)}\ket{\psi (t, \tau)}$ and $\xi_{q}=\prod_{q}^{N} \braket{\chi_q}{0_q}$. The state $\ket{\Phi_{(\text{IR},q)} (t, \tau)}$ is a massively entangled state between all field modes ~\cite{stammer_high_2022,stammer_theory_2022}. By projecting $\ket{\Phi_{(\text{IR},q)} (t, \tau)}$ onto  the state in which the harmonics are found, we project the $\ket{\Phi_{(\text{IR},q)} (t, \tau)}$ of Eq.\eqref{eq:PhiState} to the harmonic coherent states $\{\ket{\chi_q}\}_{q\geq2}$ and we obtain an IR GCSS of the form,
\begin{equation} \label{eq:cat_final}
\begin{aligned}
\ket{\text{GCSS} (t, \tau)}= \ket{\psi (t, \tau)} - \xi_{\text{IR}}(t,\tau) \ket{\alpha (t) }.
\end{aligned}
\end{equation}
The $S_{2\omega}(\tau)$ traces produced by the GCSS of Eq.\eqref{eq:cat_final} (SM \cite{SM} Pt.7) show distinct differences (Fig.~2b, c) compared to that of a coherent state (Fig.~2a).  They depict prominent beating features (black arrows in Figs.~2b and 2c) around the center  ($-20 \ \text{fs} < \tau < +20 \ \text{fs}$) and the tails ($\tau \approx \pm 45 \ \text{fs}$) of the trace. These are associated to pure non-classical effects which arise from the quantum interference between the two states composing the GCSS of Eq. \eqref{eq:cat_final}. Because the conditioning takes place upon the $2\omega$ generation in the BBO crystal, the presence of decoherence is considered negligible, therefore circumventing the decoherence problems of large amplitude GCSS. The traces shown in Figs.~2b, c have been calculated for the values of $|\delta\alpha|=1.44$ (Fig.~2b) and $|\delta\alpha|=0.24$ (Fig.~2c), which fit the experimental results shown in Figs.~2d--f . The red lines in Figs.~2a--c result from the cycle average of the 2-AC traces and they correspond to the second order intensity autocorrelation (2-IAC) traces. For both values of $|\delta\alpha|$ the value of $S_{2\omega} (\tau \approx 0)$ is significantly smaller than the corresponding value of the conventional 2-AC trace of the coherent state (Fig.~2a). For $|\delta\alpha|=0.24$ the trace (Fig.~2c) depicts a strong beating structure around the center. When $|\delta\alpha|$ is increased ($|\delta\alpha|=1.44$) (Fig.~2b), the value of $S_{2\omega} (\tau \approx 0)$ is increased resulting to a reduction of modulation depth of the beating structure around the center and the formation of a plateau--like structure.  

To visualize the quantum character of the light states leading to the beating structure of the 2-AC traces, we have calculated  in phase space $(x, p)$ their Wigner functions $W_{\omega}(x, p)$ using the values of $|\alpha|$ and $|\delta\alpha|$ used in Figs.~2a--c. $\it x$, $\it p$ are the values of the non-commuting quadrature field operators $\hat{x}=(\hat{a}+\hat{a}^\dagger)/\sqrt{2}$ and $\hat{p}=(\hat{a}-\hat{a}^\dagger)/i\sqrt{2}$, which are the analogues of the position and momentum operators of a particle in an harmonic oscillator and $\hat{a}$, $\hat{a}^\dagger$ are the photon annihilation and creation operators, respectively. For a coherent state, $W_{\omega}(x, p)$ depicts a Gaussian distribution (Fig.~2g) and leads to the conventional 2-AC trace of Fig.~2a. In case of ``medium'' (Fig.~2h) and ``small'' (Fig.~2i) GCSS the presence of quantum interference results in a non--Gaussian $W_{\omega}(x, p)$ with a negative minimum at the center and causes the beating structure in the measured 2-AC of Figs.~2e, f.

To experimentally implement the conditioning process leading to $\ket{\text{GCSS} (t, \tau)}$ of Eq.~\eqref{eq:cat_final}, we use the quantum spectrometer (QS) approach ((SM \cite{SM} Pt.3,5)) ~\cite{Tsatafyllis_2017, lewenstein_generation_2021,rivera-dean_strong_2022,stammer_high_2022,stammer_theory_2022,stammer_quantum_2023,Moiseyev_arXiv_2024, Stammer_arXiv_2024, Stammer_arXivEnergy_2024} (Fig.~1). With the QS we post select on HHG to exclude unwanted nonlinear processes.  The approach is based on shot-to-shot correlation measurements between the harmonic photocurrent signal S$_{\text{HH}}$ (integrated over $q\geq11$) recorded by the PD$_{\text{HH}}$  detector and the IR signal S$_{\text{IR}}$ recorded by the PD$_{\text{IR}}$ detector. The S$_{\text{IR}}$ and S$_{\text{HH}}$, are described by the corresponding photon number operators $\hat{I}_{\text{IR}} = \hat{a}^\dagger\hat{a}$ and $\hat{I}_{\text{HH}}= \sum_{q} \hat{b}_{q}^\dagger\hat{b}_{q}$.  When the QS is ``on'', the conditioning to HHG is achieved by selecting only the S$_{\text{IR}}$ events associated to the HHG process and effectively eliminating all residual processes.  To reveal these points, we take advantage of the energy conservation (when S$_{\text{HH}}$ increases S$_{\text{IR}}$ decreases), and we collect only those lying along the anticorrelation diagonal of the (S$_{\text{IR}}$, S$_{\text{HH}}$) joint distribution. This is the correct way to select the points where the depletion $\delta\alpha$ of the IR state leads to a shift $\chi_{q}$ on the harmonic modes ~\cite{Tsatafyllis_2017, Stammer_arXiv_2024, Stammer_arXivEnergy_2024}. In this way we effectively apply the $\hat{P}_{\text{HHG}}$ on $\ket{\psi_{(\text{IR},q)}}$ as we select those events where $\delta\alpha$ of the IR $\ket{\psi (t, \tau)}$ state is anticorrelated to the $\{\chi_{q}\}_{q\geq 2}$ shifts of the harmonic states $\{\ket{\chi_{q}}\}_{q\geq 2}$. Therefore, the IR states producing the $2\omega$ field in the BBO crystal are $\ket{\text{GCSS}}$. As a consequence of the post selection process, the success probability is reduced. In previous our works ~\cite{lewenstein_generation_2021,rivera-dean_strong_2022, stammer_quantum_2023}, the typical values of success probability which provide a GCSS with fidelity $>60\%$ was in the range of $0. 1\%$ to $0. 5\%$. In the present work the GCSS success probability is $\approx0.3\%$. The difference in the success probabilities is associated with the influence of the unwanted non-linear processes in the HHG vapor under the post selection conditions used in the present work. When the QS is ``off'',  the projection operator $\hat{P}_{\text{HHG}}$ is not applied and the IR state $\ket{\psi (t, \tau)}$ in the BBO crystal remains coherent.

The theoretical results shown in Fig.~2a--c have been confirmed experimentally. We have recorded the 2-AC trace for two values of $\delta\alpha$, which correspond to the maximum and minimum values achievable in the present experiment. The minimum $\delta\alpha$  was $\approx 6$ times lower than the maximum. The value of $\delta\alpha$ was controlled by varying the atomic density in the HHG region (SM \cite{SM} Pt.2). As is expected, when no conditioning is applied, the recorded 2-AC traces  depict the well--known features of an 2-AC trace of a pulsed coherent state. For reasons of simplicity, Fig.~2d shows only the trace recorded for the minimum value of  $\delta\alpha$. The corresponding trace for the maximum value of $\delta\alpha$ is shown in SM \cite{SM}, Pt.8.  Their shape changes drastically when we condition on HHG (Figs.~2e, f). The $S_{2\omega}(\tau)$ traces show prominent beating features (black arrows in Figs.~2e, f) around the center and at the tails of the trace. The red yellow--filled lines in Fig.~2d--f correspond to the 2-IAC traces. 

To quantitatively characterize the quantum interference effects, we have compared the $S_{2\omega} (\tau \approx 0)$ (red crossed-circles in Fig.~3a) and the modulation depth
$M=2(S_{2\omega}^{max}-S_{2\omega}^{min})/(S_{2\omega}^{max}+S_{2\omega}^{min})$ obtained from the measured 2-IAC traces (blue crossed-triangles in Fig.~3a) with the corresponding theoretical values obtained from the calculated 2-IAC traces (red line-dots and blue line-triangles in Fig.~3a). $S_{2\omega}^{max}$ and $S_{2\omega}^{min}$ are the maxima and minima of the 2-IAC traces around $\tau \approx \pm 20$ fs and $\tau \approx 0$ fs, respectively. For the coherent state, $S_{2\omega} (\tau \approx 0)=1$ and $M$ is set to zero due to the absence of modulation. The results shown in Fig.~3a further justify that the measured 2-AC traces shown in Figs.~2e, f result from driving the $2\omega$ generation with high photon number ($|\alpha| \approx 12$) ``medium'' and ``small'' GCSS with $|\delta\alpha| \approx 1.44$ and $\approx 0.24$, respectively. 
As noted in refs. ~\cite{Stammer_arXiv_2024, Stammer_arXivEnergy_2024}, deviations between experimental and theoretical results arise partly from conditioning measurements, which depend on post-selection quality and detection efficiency. It has been shown, that depending on these conditioning measurements, the fidelity of the GCSS can range from $\approx60\%$ to as high as $\approx 99.9\%$ for ideal conditional measurements. However, due to the high photon number of the GCSS used in this work, the fidelity of the state cannot be determined because of limitations in conventional approaches for assessing the density matrix of high-photon-number light states. 

It is noted that a classical mixture of coherent states does not lead to a 2-AC trace with beating structure (SM \cite{SM} Pt.9). Although the contribution of such mixtures or the presence of decoherence effects cannot be entirely excluded from the experimental measurements, these effects have not been considered in the present analysis as the main features observed in Figs.~2e, f are governed by the quantum interference between the coherent states forming the GCSS. 

To this end, we note that the method is applicable for GCSS with even higher photon numbers. This is shown in Fig.3b, where the values of $S_{2\omega} (\tau \approx 0)$ and $M$ have been calculated for $|\alpha|=30$, corresponding to an GCSS of $\approx900$ photons. The deviation from the coherent state is visible for $|\delta\alpha|<0.3$. However, in this case the quantum interference effects of the GCSS in the AC traces are less pronounced when compared to the GCSS with $|\alpha|=12$. For this reason, the present work has been conducted using GCSS with $|\alpha| \approx 12$.

\begin{figure}
\centering
    \includegraphics[width=1 \columnwidth]{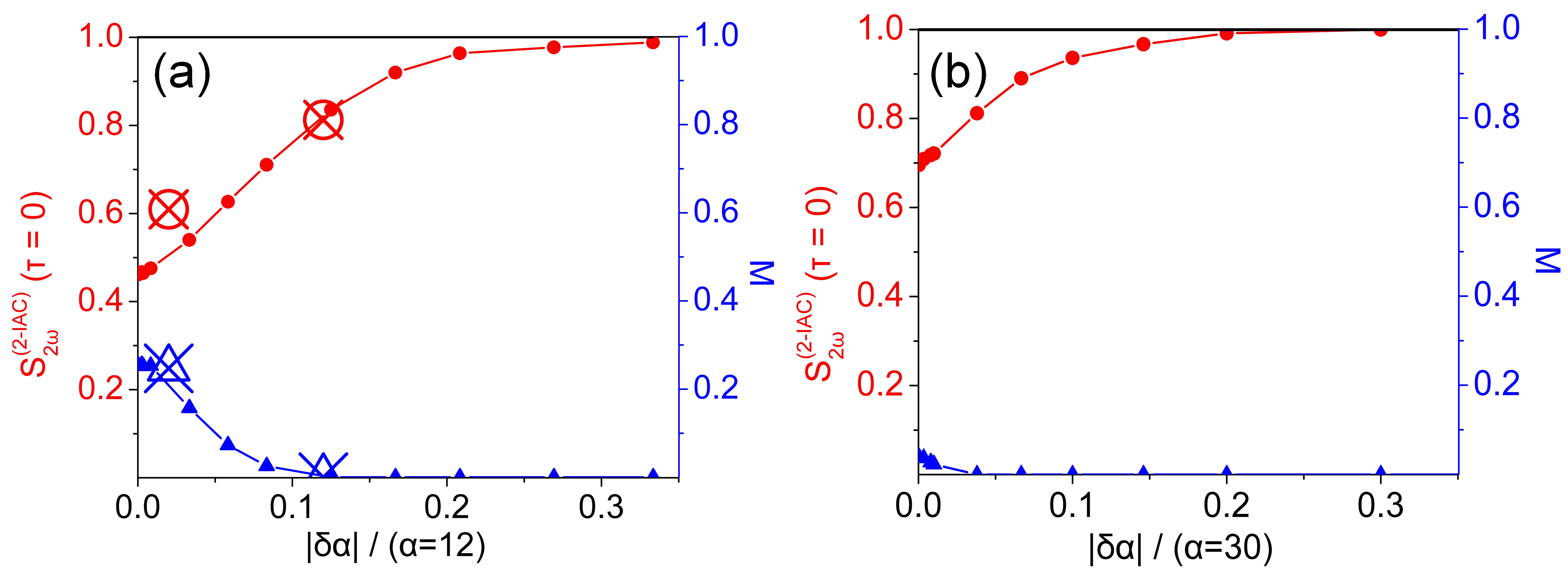}
    \caption{(a) The $S_{2\omega} (\tau \approx 0)$ (red crossed-circles) and the modulation depth $M$ (blue crossed-triangles) obtained from the measured 2-IAC traces. The size of the crossed--circles and crossed--triangles represents one standard deviation from the mean. The corresponding theoretical values are shown with the red line--dots and blue line--triangles. For the coherent state, $S_{2\omega} (\tau \approx 0)=1$ and $M=0$.  (b) Theoretically calculated values of $S_{2\omega} (\tau \approx 0)$ (red line--dots) and $M$ (blue line--triangles) for $|\alpha|=30$.}
  \label{fig:exp:main}
	\end{figure}

A direct consequence of the approach in quantum light engineering is the generation of quantum light states in different spectral regions using a nonlinear up--conversion process. We theoretically show that the spectral transfer of quantum features of light using nonlinear optics. To do this, we consider quantum light engineering by using quantum light to drive the process. Fig.4 shows the theoretically calculated (SM \cite{SM} Pt.10) Wigner functions of the $2\omega$ ($W_{2\omega}(x, p)$) when it is produced by an IR coherent state with $|\alpha| = 12$ (Fig. 4a), an IR ``medium'' GCSS with $|\alpha| = 12$ and $|\delta\alpha| = 1.44$ (Fig. 4b), and an IR ``small'' GCSS with $|\alpha| = 12$ and $|\delta\alpha| = 0.24$ (Fig. 4c). The $W_{2\omega}(x, p)$ have been calculated using Eq.\eqref{eq:InterfState} and Eq.\eqref{eq:cat_final} at $\tau=0$.  As expected, when no conditioning is applied and the state of the IR driving field is coherent, the state of generated $2\omega$ is also coherent (Fig. 4a). The situation changes when conditioning is applied and the IR driving field is a ``medium'' or ``small'' GCSS. In these cases the generated $2\omega$ depicts non--classical features which can be controlled by changing $|\delta\alpha|$. For $|\delta\alpha| = 1.44$ and $0.24$ the $W_{2\omega}(x, p)$ depict the characteristic features of a ``medium'' GCSS with small negative values (Fig. 4b), and a ``small'' GCSS with well pronounced negative values (Fig. 4c), respectively.

\begin{figure}
\centering
    \includegraphics[width=1 \columnwidth]{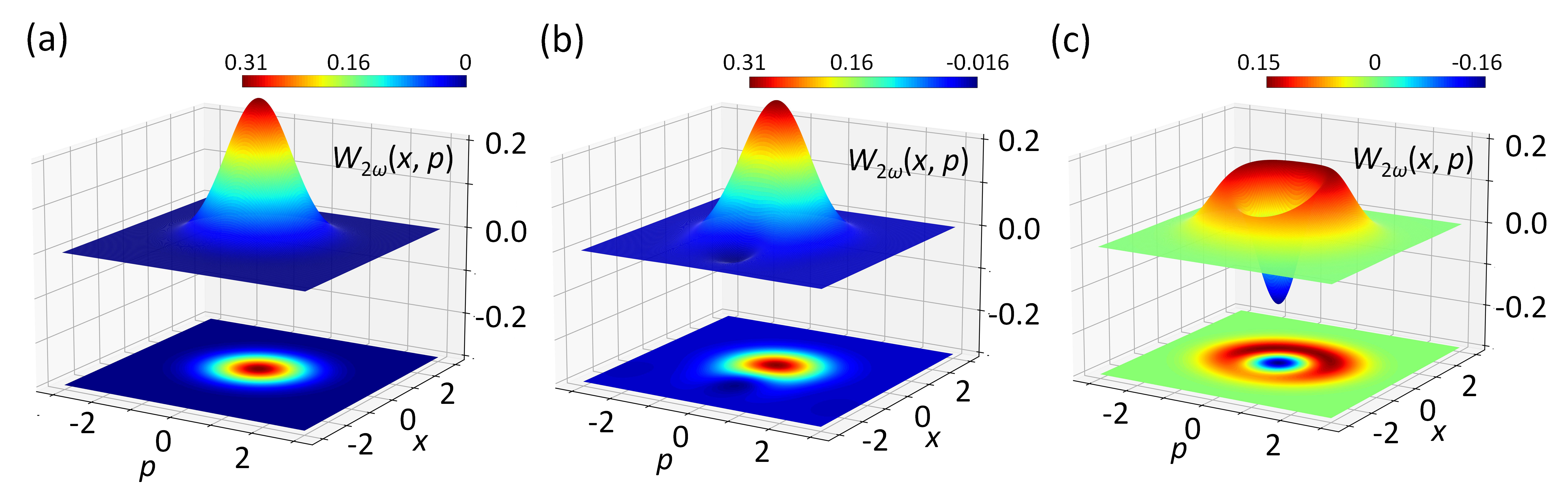}
    \caption{Theoretically calculated Wigner functions $W_{2\omega}(x, p)$ of the optical GCSS of the generated second harmonic. $W_{2\omega}(x, p)$ in phase space $(x, p)$, produced in a BBO crystal when the second harmonic generation process is driven by a coherent IR state (QS off) with $|\alpha| = 12$ (a), an IR ``medium'' GCSS (QS on) with $|\alpha| = 12$ and $|\delta\alpha| = 1.44$ (b), and an IR ``small'' GCSS (QS on) with $|\alpha| = 12$ and $|\delta\alpha| = 0.24$ (c). For clarity, the $W_{2\omega}(x, p)$ have been centered around the origin.}
        \label{fig:theory2omega:main}
	\end{figure}

In conclusion, we have created a femtosecond duration optical GCSS in the infrared spectral region with intensities sufficient to induce nonlinear processes in matter.  The non--classical properties of the light field have been characterized by means of second order autocorrelation measurements. We also theoretically show how intense infrared GCSS can be used for the generation of GCSS in the blue spectral region through a second order up-conversion process in an optical crystal. High photon number GCSS, with intensities capable of inducing nonlinear processes in matter, can be regarded as a distinctive and unparalleled resource for advancing new quantum technologies. They can be used in quantum light engineering and for the generation of massively entangled states \cite{gilchrist_schrodinger_2004,stammer_high_2022,stammer_quantum_2023, stammer2024entanglement} facilitating novel investigations in quantum information science with promising applications in quantum metrology \cite{martos2023metrological}. Furthermore, using them as drivers for nonlinear optical processes, opens up new possibilities in various research fields in quantum information science including quantum sensing \cite{giovannetti_quantum-enhanced_2004,gilchrist_schrodinger_2004,giovannetti_advances_2011,joo_quantum_2011, martos2023metrological}, nonlinear spectroscopy \cite{dorfman_nonlinear_2016} and quantum imaging \cite{moreau_imaging_2019}. The findings are of broad importance, as they pave the way for a novel quantum nonlinear spectroscopy based on the interplay between the quantum properties of light and those of quantum matter on ultrafast timescales.  \\

{\it Acknowledgments:} The Hellenic Foundation for Research and Innovation (HFRI) and the General Secretariat for Research and Technology (GSRT) under grant agreement CO2toO2 Nr.:015922, the European Union’s HORIZON-MSCA-2023-DN-01 project QU-ATTO under the Marie Skłodowska-Curie grant agreement No 101168628, the LASERLABEUROPE V (H2020-EU.1.4.1.2 grant no.871124) and ELI--ALPS. ELI--ALPS is supported by the EU and co-financed by the European Regional Development Fund (GINOP No. 2.3.6-15-2015-00001). H2020-EU research and innovation program under the Marie Skłodowska-Curie (No. 847517). Government of Spain (Severo Ochoa CEX2019-000910-S and TRANQI), Fundació Cellex, Fundació Mir-Puig, Generalitat de Catalunya (CERCA program) and the ERC AdG CERQUTE. ERC AdG NOQIA; MCIN/AEI (PGC2018-0910.13039/501100011033, CEX2019-000910-S/10.13039/501100011033, Plan National FIDEUA PID2019-106901GB-I00, STAMEENA PID2022-139099NB, I00, project funded by MCIN/AEI/10.13039/501100011033 and by the EU Next Generation EU/PRTR (PRTRC17.I1), FPI). QUANTERA MAQS PCI2019-111828-2; QUANTERA DYNAMITE PCI2022-132919, QuantERA II Programme co-funded by H2020-EU program (No 101017733); Ministry for Digital Transformation and of Civil Service of the Spanish Government through the QUANTUM ENIA project call-Quantum Spain project, and by the EU through the Recovery, Transformation and Resilience Plan—Next Generation EU within the framework of the Digital Spain 2026 Agenda; Fundació Cellex; Fundació Mir-Puig; Generalitat de Catalunya (European Social Fund FEDER and CERCA program, AGAUR Grant No. 2021 SGR 01452, QuantumCAT \textbackslash{} U16-011424, co-funded by ERDF Operational Program of Catalonia 2014-2020); Barcelona Supercomputing Center MareNostrum (FI-2023-1-0013) funded by the EU. Views and opinions expressed are however those of the author(s) only and do not necessarily reflect those of the EU, European Commission, European Climate, Infrastructure and Environment Executive Agency (CINEA), or any other granting authority. Neither the EU nor any granting authority can be held responsible for them (EU Quantum Flagship PASQuanS2.1, 101113690, EU H2020 FET-OPEN OPTOlogic (No 899794), EU Horizon Europe Program (No 101080086-NeQST); ICFO Internal “QuantumGaudi” project; EU H2020 program under the Marie Sklodowska-Curie (No. 847648); “La Caixa” Junior Leaders fellowships, La Caixa” Foundation (ID 100010434): CF/BQ/PR23/11980043. P.S. acknowledges funding from the European Union’s Horizon 2020 research and innovation programme under the Marie Skłodowska-Curie grant agreement No 847517.

{\it Author contributions}: Th. L, J. R--D, P. S. equally contributed authors. Th. L. contributed to the experimental runs and data analysis. J. R--D and P. S. contributed to the development of the theoretical approach and data analysis. M. L.: Supervised the theoretical part of the work. P.T. conceived and supervised the project. 

\bibliography{References_PRL}{}

\begin{thebibliography}{47}
\bibitem [{not()}]{note_delay}%
  \BibitemOpen
  \href@noop {} {}\bibinfo {note} {Under this framework, a time-delay of $\tau$ would affect the coherent state amplitude as $\alpha_{L}f(t)e^{i \omega t} \to \alpha_{L}f(t+\tau)e^{i \omega (t+\tau)}$, leading to $\langle \hat{E}\rangle \propto \lvert\alpha_L\rvert f(t+\tau)\sin(\omega(t+\tau)+\phi)$.}\BibitemShut {Stop}%
\end{thebibliography}

\end{document}


\title{Supplementary Material for:\\ Nonlinear optics using intense optical coherent state superpositions}

\author{Th. Lamprou}
 \affiliation{Foundation for Research and Technology-Hellas, Institute of Electronic Structure \& Laser, GR-7001 Heraklion (Crete), Greece}

\author{J.~Rivera-Dean}
\affiliation{ICFO -- Institut de Ciencies Fotoniques, The Barcelona Institute of Science and Technology, 08860 Castelldefels (Barcelona)}

\author{P. Stammer}
\affiliation{ICFO -- Institut de Ciencies Fotoniques, The Barcelona Institute of Science and Technology, 08860 Castelldefels (Barcelona)}
\affiliation{Atominstitut, Technische Universit\"{a}t Wien, 1020 Vienna, Austria}

\author{M. Lewenstein}
\affiliation{ICFO -- Institut de Ciencies Fotoniques, The Barcelona Institute of Science and Technology, 08860 Castelldefels (Barcelona)}
\affiliation{ICREA, Pg. Llu\'{\i}s Companys 23, 08010 Barcelona, Spain}

\author{P. Tzallas}
\email{ptzallas@iesl.forth.gr}
\affiliation{Foundation for Research and Technology-Hellas, Institute of Electronic Structure \& Laser, GR-7001 Heraklion (Crete), Greece}
\affiliation{ELI-ALPS, ELI-Hu Non-Profit Ltd., Dugonics tér 13, H-6720 Szeged, Hungary}

\maketitle

\noindent\textbf{{Part 1: Theoretical description}}\\

According to the fully quantized theory of high harmonic generation (HHG) ~\cite{lewenstein_generation_2021,rivera-dean_strong_2022,stammer_high_2022,stammer_theory_2022, stammer_quantum_2023} for a single atom scenario, before the interaction (Fig. S1a,b) the IR driving laser field, the harmonic modes and the atom are described by the product state $\ket{\Psi(t_{0})} = \ket{g} \otimes \ket{\alpha_{L}} \bigotimes_{q \geq 2} \ket{0_{q}}$. Here, $\ket{g}$ is the ground state of the atom, and $\ket{\alpha_{L}}$, $\ket{0_q}$ are the coherent states of the driving field and the vacuum state of the $q$th harmonic, respectively. In the following, we denote $q=1\equiv L$.
\par
The Hamiltonian characterizing the dynamics between the laser and the atom within the single-active electron approximation is given by $\hat{H} = \hat{H}_{a} +\hat{H}_{f}+\hat{H}_{int}$. Here, $\hat{H}_{a}$ and $\hat{H}_f=\sum_{q} \hbar\omega_{q}\hat{a}_q^\dagger \hat{a}_q$ are the atomic and free-electromagnetic field Hamiltonians, respectively, with $\hat{a}_q^\dagger$($\hat{a}_q$) the creation (annihilation) operator acting on the field mode with frequency $\omega_{q}$ ($q=1$ corresponds to the fundamental IR driving field). $\hat{H}_{int}$ describes the light-matter interaction which, in the length gauge and under the dipole approximation, is given as $\hat{H}_{int} = \hat{\mathbf{d}}\cdot \hat{\mathbf{E}}$, with $\hat{\mathbf{d}}$ the dipole moment operator and $\hat{\mathbf{E}} = i\mathbf{g}(\omega_{L})\sum_{q=1}(\hat{a}_{q} - \hat{a}^{\dag}_{q})$ the electric field operator. After applying a set of unitary transformations in order to simplify the description of the dynamics, the Schr\"{o}dinger equation reads,

\renewcommand{\theequation}{S.\arabic{equation}}
\begin{equation}\label{SM_Theory_Eq:SE_interaction}
    i \hbar \dv{t} \ket{\psi(t)} = \hat{\mathbf{d}}(t) \cdot \hat{\mathbf{E}}(t) \ket{\psi(t)},
\end{equation}
where the initial condition is now given by $\ket{\psi(t_{0})} = \ket{g}\bigotimes_{q=1}\ket{0_{q}}$. It is worth noting that the fundamental mode appears in a vacuum state due to the presence of a displacement of our frame of reference by $\alpha_{L}$ among the aforementioned unitary transformations. In order to study the process of HHG we need to condition the interaction of the atomic ground state, i.e. project Eq. (\ref{SM_Theory_Eq:SE_interaction}) onto $\ket{g}$. Furthermore, we denote the state of light as $\ket{\Phi(t)} = \braket{g}{\psi(t)}$. Under the strong field approximation, the dynamics of the HHG process is described by

\begin{equation}\label{SM_Theory_Eq:SE_effective_Hamiltonian}
    i\hbar \dv{t} \ket{\Phi(t)} \approx \expval{\mathbf{d}(t)} \cdot \hat{\mathbf{E}}(t) \ket{\Phi(t)},
\end{equation}
where $\expval{\mathbf{d}(t)} = \mel{g}{\hat{\mathbf{d}}(t)}{g}$ is the time-dependent expectation value of the dipole moment evaluated over the atomic ground state. The effective Hammiltonian in Eq. (\ref{SM_Theory_Eq:SE_effective_Hamiltonian}) is a linear form of photon creation and annihilation operators. Thus, the unitary evolution operator is an exponent of a linear form of creation and annihilation operators, and thus when acting on coherent states, it will shift them. In other words, the interaction leading to HHG can be described by the following multi-mode displacement operator (after returning to the original frame of reference) $\hat{D}(\chi) = \prod_{q=1} \hat{D}(\alpha_L \delta_{q,1}+\chi_{q})$, where $\delta_{q,1}$ is the Kronecker-Delta and $\chi_{q} = -i\kappa \sqrt{q} \expval{d}(q\omega)$, with coupling constant $\kappa$, and the Fourier transform of the time-dependent dipole moment expectation value $\expval{d}(q\omega) = \int_{-\infty}^{\infty} \dd{t'} \expval{d(t)}e^{iq \omega t}$. Thus, the optical state after the interaction is,

\begin{equation}
    \ket{\phi} = \bigotimes_{q=1} \hat{D}(\chi_{q})\ket{0_{q}}
\end{equation}

\par
In the present work, we consider the following quantum optical (fundamental and harmonics) state before the interaction $\ket{\phi_{\text{I}}(t)} = \ket{\alpha_{L}(t)}\bigotimes_{q=2}\ket{0_{q}}$, where $\alpha_{L}(t) = \alpha_{L}f(t)e^{i \omega t}$, with $\alpha_{L},f(t),\omega$ account for the amplitude, envelope and central frequency of the IR pulse, respectively. In the following, we effectively account for the pulse envelope within the coherent state amplitude in order to properly account for the time-delays in interferometric setups within single mode descriptions of the IR laser pulse\cite{note_delay}. Then, the multi-mode displacement operator, reads $\hat{D}(\delta \alpha_{L})\prod_{q\geq 2}\hat{D}(\chi_{q})$ (Fig. S1b). The shift of the fundamental mode ($\delta \alpha_{L}$) captures the IR depletion due to HHG process which, consequently induces a displacement $\chi_{q}$ to the generated harmonic modes. When acting on $\ket{\phi_{\text{I}}(t)}$ we end up with the final light state after the interaction
\begin{equation} \label{SM_Theory_Eq:CohAfterHHG}
    \ket{\phi(t)} = \ket{(\alpha_{L}+\delta \alpha_{L})f(t) e^{i\omega t}}\bigotimes_{q} \ket{\chi_{q}}.
\end{equation}

\par 
As the HHG process drives the displacement of all modes, it is important to note that the depletion of the IR driving field and the shift of the harmonic modes are inherently linked. Thus, the mode that is actually excited during the HHG process is a wavepacket mode (described by a set of multimode states $\{\ket{\Tilde{n}}\}$) taking into account these correlations (Fig. S1b). Here, the state $\ket{\Tilde{0}}$ describes the case where no harmonic radiation is generated, and therefore corresponds to the initial quantum optical state prior to the HHG interaction. In contrast, $\ket{\Tilde{n}}$ with $\Tilde{n}\geq 0$, describes the case where harmonic radiation is generated. In other words, the excitation of the wave packet $\ket{\Tilde{n}}$ expresses that the ``creation'' of energy in the harmonic modes requires the ``annihilation'' of energy from the IR field, reflecting the up-conversion process of the IR photons towards HHG. This allows us to define a set of positive operators $\{\hat{\Pi}_{\Tilde{0}}, \hat{\Pi}_{\Tilde{n}\neq \Tilde{0}}\}$, describing when the HHG process and the corresponding IR depletion have occurred or not, respectively, or equivalently whether the wave packet (which takes into account the correlations between the field modes) has been excited or not. Specifically, the element $\hat{\Pi}_{\Tilde{0}} = \dyad{\Tilde{0}}$ projects onto the subspace where no excitations are found, while $\hat{\Pi}_{\Tilde{n}\neq \Tilde{0}} = \sum_{\Tilde{n}\neq \Tilde{0}}  \dyad{\Tilde{n}} = 1-\dyad{\Tilde{0}} = \hat{P}_{\text{HHG}}$ onto the subspace where HHG excitations are found. 

We refer to $\hat{P}_{\text{HHG}} \equiv \mathbbm{1}-\dyad{\Tilde{0}}$ as the conditioning on HHG (Fig. S1c). As has been described in the main text, the act of this operator leads to the $\ket{\text{GCSS}}$ state of Eq. (3) of the main text of the manuscript. As is shown in ref. \cite{Stammer_arXiv_2024, Stammer_arXivEnergy_2024}, the projection $\mathbbm{1}-\dyad{\alpha}$ sufficiently approximates the exact action of the conditioning operator. 
\\

\noindent\textbf{{Part 2: Experimental approach}}\\
For reasons of readability of the section, in Fig.S1a we show the optical layout of the experimental arrangement shown in Fig.1 of the main text of the manuscript. This figure includes the beam separator BS$_1$ which reflects a small portion of the IR beam towards the IR photodetector PD$_0$. This configuration has been used to measure in each shot the energy (photon number) of the driving field before HHG. Only the laser shots with intensity fluctuations $<0.5$\% of the mean have been used. Also, includes the Al filter, and the prism which allows the detection of the harmonics with order $q\geq 11$ and the $2\omega$ frequency generated in the BBO crystal, respectively. For reasons of simplicity, these components have not been included in Fig.1 of the main text of the manuscript. The Fig.S2a  shows the IR power spectrum after the BBO crystal. Fig. S2b shows the spectrum of the harmonics reaching the XUV photodetector PD$_{\text{HH}}$.
\\

The whole system was operating at 0.5 kHz repetition rate collecting the data for each laser shot. The experiment was performed using a linearly polarized $\approx 25$ fs IR laser pulse of $\lambda=800$ nm. The laser pulse was focused with an intensity $\sim 10^{14}$ W per cm$^{2}$ into Argon atoms where odd harmonics of frequencies ($q\omega$ and $q\leq35$) in the extreme-ultraviolet (XUV) spectral range were generated (Fig. S2b). The Argon atoms introduced in the interaction region by means of piezo--based pulse gas jet (Fig. S1a). The harmonics were separated by the IR field by means of a multilayer infrared-antireflection coating plane mirror (harmonic separator HS) placed at grazing incidence angle. This allows the fundamental field to pass through and reflects the harmonics towards an XUV photodetector PD$_{\text{HH}}$. An $\approx 150$ nm thick Aluminum filter was placed before the PD$_{\text{HH}}$ (Fig. S1a). This filter allows only the harmonics with order $q\geq 11$ to pass through (Fig. S2b). After the HS the beam was passing through an energy attenuator F,  used to control the energy of the IR beam. In F, for reasons of simplicity, we include all the energy losses introduced by the optical elements after the HHG. The state of the IR laser field before the laser--atom interaction is a coherent state $\ket{\alpha_L (t)}$, after the interaction is an amplitude depleted coherent state $\ket{(\alpha_L+\delta\alpha_L) (t)}$ and after F is an attenuated coherent state $\ket{(\alpha+\delta\alpha) (t)}$. The change of $\delta\alpha$ was achieved by varying the atomic density in the HHG region ~\cite{rivera-dean_strong_2022}. Since $\delta\alpha \propto N_{\text{at}}$ (where $N_{\text{at}}$ is the number of atoms participating in the HHG process), its change has been traced by measuring the harmonic yield ($Y$) which is $Y \propto N_{\text{at}}^2$ and thus $|\delta\alpha| \propto Y^{1/2}$. Then, the beam passes through a Mach-Zehnder interferometer which includes a phase shifter $\varphi=\omega\tau$ in one of its arms. The beams exiting the interferometer passes through a beam splitter (BS) which reflects the 50 $\%$ of the energy of both IR beams towards an IR photodetector PD$_{\text{IR}}$. Because PD$_{\text{IR}}$ measures the energy of both IR beams exiting the interferometer, its signal is invariant with $\tau$.  All photodetectors are operating in the linear regime. \\

\noindent\textbf{{Part 3: Conditioning on HHG using QS }}\\
The signals of PD$_{\text{0}}$, PD$_{\text{IR}}$ and PD$_{\text{HH}}$ have been used by the quantum spectrometer (QS) approach ~\cite{Tsatafyllis_2017, lewenstein_generation_2021,rivera-dean_strong_2022,stammer_high_2022,stammer_theory_2022,stammer_quantum_2023, Moiseyev_arXiv_2024, Stammer_arXiv_2024, Stammer_arXivEnergy_2024}.  The aim of QS is to select the S$_{\text{IR}}$ events associated only with the HHG process. Specifically, the QS is based on shot-to-shot correlation measurements between the harmonic photocurrent signal S$_{\text{HH}}$ (integrated over $q\geq11$) recorded by the PD$_{\text{HH}}$  detector (blue points in Fig.S1c) and the IR signal S$_{\text{IR}}$ recorded by the PD$_{\text{IR}}$ detector (black points in Fig.S1c). The gray points in Fig.S1c show the (S$_{\text{IR}}$, S$_{\text{HH}}$) joint distribution. The S$_{\text{IR}}$ and S$_{\text{HH}}$, are described by the corresponding photon number operators $\hat{I}_{\text{IR}} = \hat{a}^\dagger\hat{a}$ and $\hat{I}_{\text{HH}}= \sum_{q} \hat{b}_{q}^\dagger\hat{b}_{q}$.  When the QS is ``on'', the conditioning to HHG is achieved by selecting only the points along the anticorrelation diagonal (green points in Fig. S1c) of the (S$_{\text{IR}}$, S$_{\text{HH}}$) joint distribution.  This is a physically acceptable way (based on energy conservation) to select the S$_{\text{IR}}$ events associated only with the HHG process. By selecting these points, we effectively apply the $\hat{P}_{\text{HHG}}$ on $\ket{\psi_{(\text{IR},q)}}$ as we select those events where $\delta\alpha$ of the IR $\ket{\psi (t, \tau)}$ state is anticorrelated to the $\{\chi_{q}\}_{q\geq 2}$ shifts of the harmonic states $\{\ket{\chi_{q}}\}_{q\geq 2}$. Therefore, the IR states producing the $2\omega$ in the BBO crystal (red points in Fig.S1c) are $\ket{\text{GCSS}}$. When the QS is ``off'',  the projection operator $\hat{P}_{\text{HHG}}$ is not applied and the IR state $\ket{\psi (t, \tau)}$ in the BBO crystal remains coherent.\\

\renewcommand{\thefigure}{S\arabic{figure}}
\begin{figure*}
    \centering
    \includegraphics[width=0.95 \textwidth]{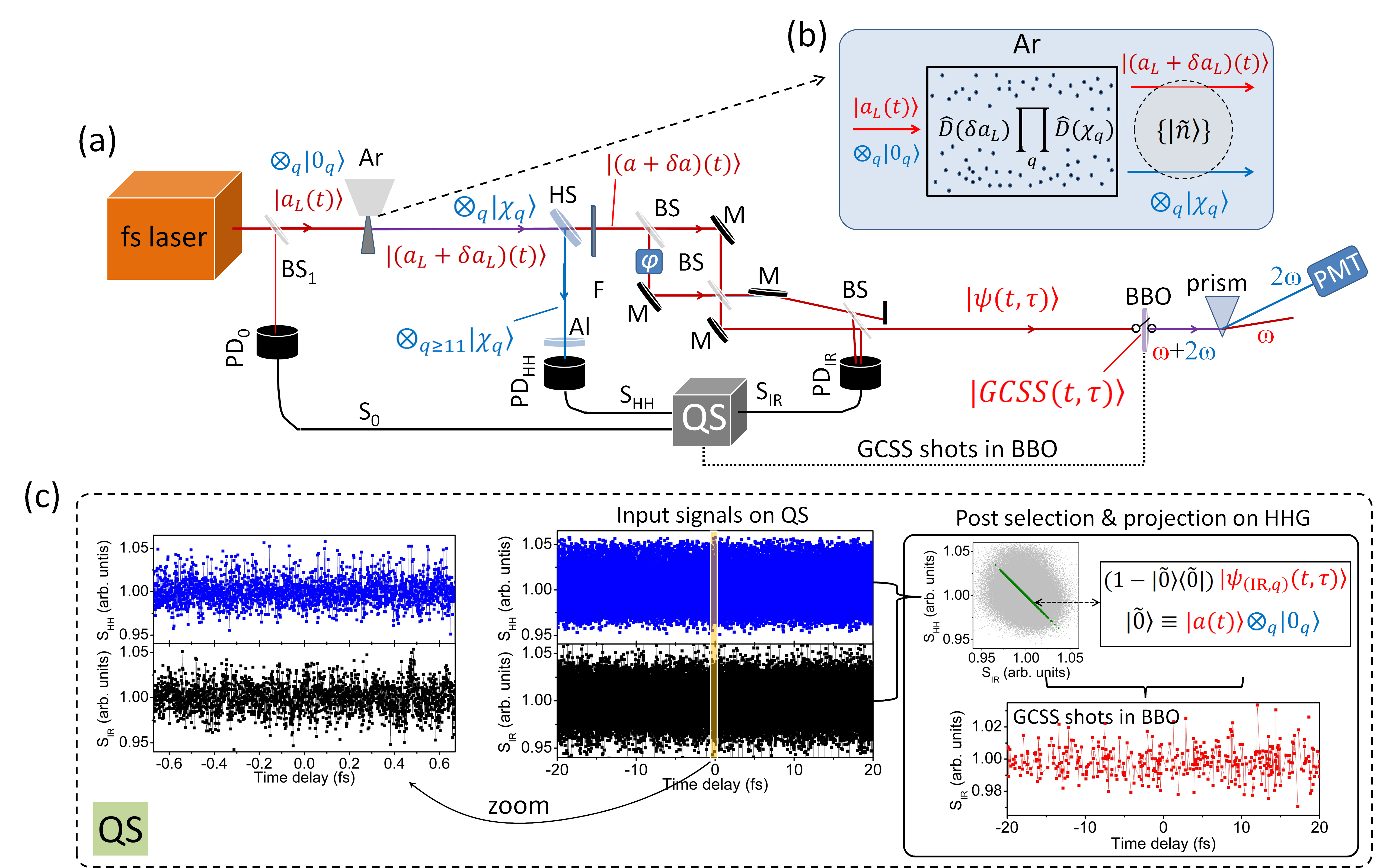}
    \caption{Optical layout of the experimental approach. (a) $\ket{\alpha_L(t)}\bigotimes_{q}\ket{0_q}$ is the initial coherent field state. $\ket{\alpha_L(t)}$ and $\ket{0_q}$ represent the state of the 25 fs IR laser pulse of frequency $\omega$ and the vacuum states $q$th harmonic, respectively.  The fs laser pulse interacts the Argon atoms and high odd-order harmonics of frequencies ($q\omega$) in the extreme-ultraviolet spectral range are generated.  $\ket{(\alpha_{L} + \delta \alpha_{L}) (t)} \bigotimes_{q}\ket{\chi_{q}}$ is the coherent state after the HHG. $\ket{(\alpha_{L} + \delta \alpha_{L}) (t)}$ and $\ket{\chi_{q}}$ are the coherent states of the depleted by the HHG process IR field and the generated $q$th harmonic, respectively. HS is a harmonic separator which reflects the high harmonics towards HH photodetector (PD$_{HH}$). $Al$ is a 150 nm thick Aluminum filter which allows only the harmonics with order $q \geq 11$ to pass through. PD$_{\text{HH}}$. HS, F are harmonic separator, IR attenuator . The state $\ket{(\alpha + \delta \alpha) (t)}$ enters the Mach-Zehnder interferometer. BS, M are beam splitters, plane mirrors. The phase shifter ($\varphi$) introduces a time delay $\tau$ between the two coherent states in the interferometer. The outgoing from the interferometer IR state $\ket{\psi (t, \tau)}$ was focused into a BBO crystal where the second harmonic ($2\omega$) is generated. A prism placed after the BBO crystal separates the second harmonic from the fundamental. The $2\omega$ photons were detected by means of a photomultiplier (PMT). (b) Laser--Argon interaction. $\ket{\alpha_L(t)}\bigotimes_{q}\ket{0_q}$ is the initial coherent field state. $\hat{D}(\delta\alpha_{L})\prod_{q}\hat{D}(\chi_{q})$ describes the multimode amplitude shifts ($\delta\alpha_{L}$ and $\chi_q$ for IR and harmonics, respectively) induced by the interaction. $\{\ket{\tilde{n}}\}$ is the excited multimode wavepacket which accounts the shift correlations. $\ket{(\alpha_{L} + \delta \alpha_{L}) (t)} \bigotimes_{q}\ket{\chi_{q}}$ is the coherent state after the HHG. (c) QS approach. The QS creates the optical GCSS $\ket{\text{GCSS}(t,\tau)}$ by projecting ($\mathbbm{1}-\dyad{\Tilde{0}}$) the $\ket{\psi_{(\text{IR},q)}}$ state on HHG. S$_{\text{HH}}$ (blue points) and S$_{\text{IR}}$ (black points) are the photocurrents recorded by the PD$_{\text{HH}}$ and PD$_{\text{IR}}$ detectors. The gray and green points show the (S$_{\text{IR}}$, S$_{\text{HH}}$) joint distribution and the anticorrelation diagonal, respectively.  The red points correspond to the $\ket{\text{GCSS}}$ states producing the $2\omega$.}
        \label{fig:SM:setup}
	\end{figure*}

\begin{figure}
    \centering
    \includegraphics[width=1 \columnwidth]{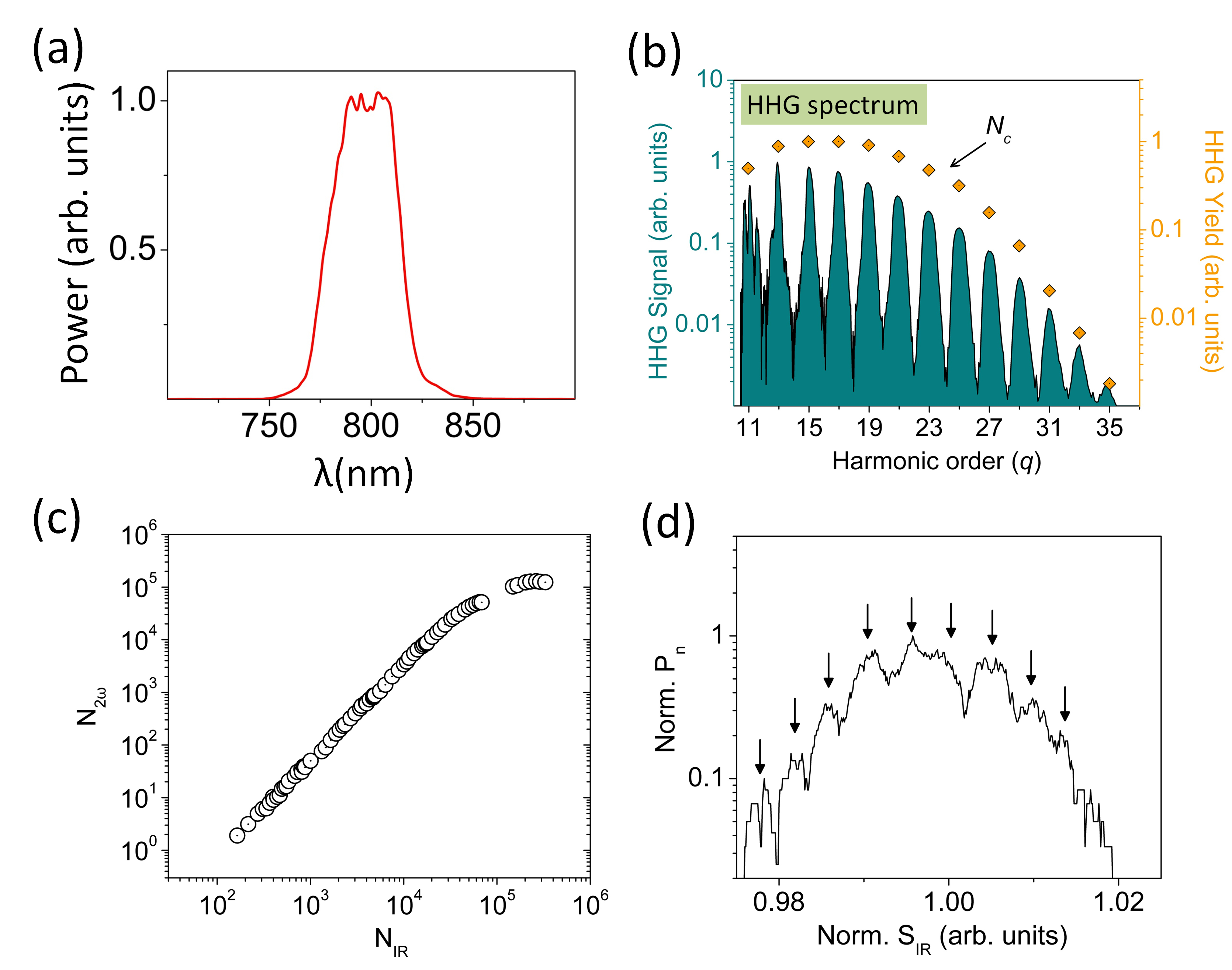}
 \caption{(a) IR power spectrum after the BBO crystal. (b) High harmonic spectrum (green filled area) generated by the interaction of Ar atoms. The orange points show the harmonic yield which corresponds to the integral of each harmonic peak. $N_c$ depicts the cut--off harmonic frequency, after which the harmonic yield drops rapidly. (c) Dependence of the $2\omega$ photons (N$_{2\omega}$) generated on the BBO crystal on the photon number of the IR field on the crystal N$_{\text{IR}}$. The deviation from the initially quadratic scaling is attributed to saturation of the $2\omega$ generation process. The calibration of N$_{2\omega}$ was according to the minimum signal of the PMT detector which corresponds to an N$_{2\omega}$ in the range of few--photons. (d)  Probability of absorbing IR photons towards the harmonic emission. }
        \label{fig:SM:QSCond}
	\end{figure}

\noindent\textbf{{Part 4: Detection of the $2\omega$ generated in the BBO crystal}}\\
Before the BBO crystal, the photon number of the IR field was reduced to $\approx 150$ photons per pulse. The beam was focused by an $f=5$ cm focal length lens in to a BBO crystal leading to the generation of few $2\omega$ photons. The intensity of the IR on the BBO crystal was $\approx 3 \times 10^4$ W per cm$^2$. A prism placed after the BBO crystal separates the second harmonic from the fundamental. The $2\omega$ photons were detected by means of a photomultiplier (PMT) of quantum efficiency $\approx 0.3$ at 400 nm. The collection efficiency of the arrangement was $\approx 1$, i.e., all the generated photons reach the detector. The dependence of the $2\omega$ photon number (N$_{2\omega}$) generated on the BBO crystal on the photon number of the IR field N$_{\text{IR}}$ (measured by means of an IR photodiode just before the crystal) is shown in Fig. S2c. The N$_{2\omega}$ was calibrated according to the minimum signal of the PMT detector which corresponds to N$_{2\omega}$ in the range of few--photons. The PMT signal ($S_{2\omega}$) was recorded for each laser shot as a function of the delay $\tau$ between the two fields.
\\

\noindent\textbf{{Part 5: QS}}\\
The aim of QS is to select the S$_{\text{IR}}$ events associated only with the HHG process. Specifically, it selects the IR photons that have been absorbed from the driving field for the HHG. The operation of QS relies on photon statistics and the shot-to-shot correlation between the photocurrents S$_{\text{HH}}$, S$_{\text{IR}}$ and the energy conservation ~\cite{lewenstein_generation_2021,rivera-dean_strong_2022,stammer_high_2022,stammer_theory_2022,stammer_quantum_2023, Tsatafyllis_2017}, i.e., when S$_{\text{HH}}$ increases, the S$_{\text{IR}}$ decreases ~\cite{Tsatafyllis_2017}. Considering that $N_{\text{IR}}$ is the IR photon number before the laser--Ar interaction (measured by PD$_0$), the photon number of the IR field and harmonics after the interactions is $N'_{\text{IR}}$ and $N_{\text{HH}}$, respectively, and due energy conservation  $N'_{\text{IR}}\leq N_{\text{IR}}$.  Due to amplitude attenuation introduced by the optical elements in the optical arrangement, the IR and HH photon numbers reaching the detectors PD$_{\text{IR}}$ and PD$_{\text{HH}}$ are $n'_{\text{IR}}$ and $n_{\text{HH}}$, respectively. These are related with $N_{\text{HH}}$ and $N'_{\text{IR}}$ through the equations $n_{\text{HH}} = N_{\text{HH}}/A_{\text{HH}}$,  and $n'_{\text{IR}} = N'_{\text{IR}}/B_{\text{IR}}$ where $A_{\text{HH}}$ and $B_{\text{IR}}$ are the attenuation factors corresponding to the HH and IR attenuation. The photon number signals measured by PD$_{\text{0}}$, PD$_{\text{IR}}$ and PD$_{\text{HH}}$, are recorded for each laser shot by a high dynamic range boxcar integrator, resulting in photocurrent outputs S$_{\text{0}}$, S$_{\text{IR}}$ and S$_{\text{HH}}$, respectively. Since the intensity dependence of the generated harmonic photons is the same with the IR photon losses ~\cite{Tsatafyllis_2017}, the variance of S$_{\text{HH}}$ is set to be balanced to the variance of S$_{\text{IR}}$.  Then, we create the joint distribution (S$_{\text{IR}}$, S$_{\text{HH}}$) shown with gray points in Fig. S1c. The distribution contains information of all processes occurring during the laser-atom interaction, and provides access to the correlated HH-IR signals. Taking into account the fact that the generation of $N_q$ photons of the $q$th harmonic corresponds to $qAN_q$ IR photons lost (where $A$ is the HH absorption factor in the HHG medium), information about the probability of absorbing IR photons towards harmonic generation can be extracted. As HHG is a small fraction compared to all processes taking place in the laser-Ar interaction region, the points that correlate the IR photon losses to the generated HH photons is a small portion of the number of points of the joint (S$_{\text{IR}}$, S$_{\text{HH}}$) distribution. To reveal these points, we take advantage of the energy conservation (when S$_{\text{HH}}$ increases S$_{\text{IR}}$ decreases), and we collect only those lying along the anticorrelation diagonal of the joint distribution (in the present experiment the number of selected points was  $\approx 0.4 \%$ of the number of points of the joint (S$_{\text{IR}}$, S$_{\text{HH}}$) distribution). This is a physically acceptable way to select the points where the depletion $\delta\alpha$ of the IR state $\ket{\psi (t, \tau)}$ leads to a shift $\chi_{q}$ of the harmonic modes.  A consequence of this selection is reflected in the multipeak structure of the IR probability distribution $P_n$ (Fig.S2d), with a spacing between the peaks to be $\propto(\Delta q)N_q$  (where $\Delta q = 2$ for the odd order harmonics). The $P_n$ provides the probability of absorbing IR photons towards the harmonic emission.\\

\noindent\textbf{Part 6: Experimental procedure and data analysis}\\
The procedure that we have followed to record the 2-AC traces shown in Figs.~2d--f is the following. We have measured for each laser shot the signals of PD$_{\text{IR}}$, PD$_{\text{HH}}$, PMT as a function of the delay $\tau$. The electronic noise has been subtracted for each laser shot from all detected signals. We have used only the laser shots with intensity fluctuations $<0.5$\% of the mean. Then, we recorded a 2-AC trace without conditioning (QS off) (conventional 2-AC trace of a coherent state). The number of shots (points in the 2-AC trace) accumulated was $\sim 5\times 10^5$ with $\approx 310$ points per $\approx 0.1$ fs. All these points are included in the joint (S$_{\text{IR}}$, S$_{\text{HH}}$) distribution shown with gray points in Fig.S1c. To obtain the traces shown in Fig.~2e (``medium'' GCSS) and Fig.~2f (``small'' GCSS), in a post selection process, we condition on the HHG (QS on) for two values of $|\delta\alpha|$, i.e., for each value of $|\delta\alpha|$ we create the joint (S$_{\text{IR}}$, S$_{\text{HH}}$) distribution and from this we select the shots along  the anticorrelation diagonal (green points in Fig.S1c). After the post selection process, the number of points remained in each of the traces (shown in Fig.~2e,f) is $\approx 1.7\times10^3$. In order to obtain the cycle average of the post selected traces we have applied a numeric band block frequency filtering process \cite{Boitier_2009, Boitier_2011}. Since after the post selection process the spacing between the points in time delay axis is not the same, in order to properly apply the band block frequency filtering process an interpolation procedure, which does not affect the traces, has been implemented. In order to have the same number of points in the conventional 2-AC trace of the coherent state shown in Fig.~2d and Fig.~S3, we have randomly selected $\approx 1.7\times10^3$ points from the $\sim 5\times 10^5$ points of the initially recorded 2-AC trace. This point reduction procedure does not have any influence on the 2-AC trace of the coherent state. This is because an $\approx 160$ fs long trace which contains $\approx 1.7\times10^3$ points results to a point per $\approx 0.1$ fs which is sufficient to resolve the $\approx 2.67$ fs cycle of the IR field.  This has been also verified by comparing the two 2-AC traces. The red yellow--filled points in Fig. 2d--f and Fig. S3, have been obtained by applying a numeric band block frequency filtering process \cite{Boitier_2009, Boitier_2011} which blocks all the frequencies $>0.2$ fs$^{-1}$ (including the $\approx 0.37$ fs$^{-1}$ frequency of the interferogram) and then averaging over $\approx 25$ points included in a cycle of the interferogram. The error bar of these points, which represents one standard deviation from the mean, has been obtained from $\sigma_{\text{(2-IAC)}}= \sigma_{\text{(2-AC)}}/\sqrt{25}$. $\sigma_{\text{(2-AC)}}$ is the standard deviation from the mean in the 2-AC traces.\\

\noindent\textbf{{Part 7: 2nd order autocorrelation for GCSS}}\\
As described in the main text of the manuscript, the IR field after recombination at the final BS of the interferometer interacts with a nonlinear crystal and generates the $2\omega$ field. The signal $S_{2\omega}$ is proportional to the squared of the intensity of the incoming IR field. Incorporating the time delay $\tau$ the $S_{2\omega}(\tau)$ reads, 
    \begin{align}
        S_{2\omega}(\tau) = \eta \int dt \expval{I^2}(t,\tau),
    \end{align}
    where $I =a^\dagger a$ is the intensity operator of the IR field. Using that $\expval{I^2} = \expval{(a^\dagger)^2a^2} + \expval{a^\dagger a}$ scales as $\expval{I^2} = \expval{n}^2 + \expval{n}$ such that we can neglect the second term in case of high average photon numbers. In our case, $\expval{n} \sim 10^2$ IR photons drive the nonlinear process of second harmonic generation. 
    First, we compute the signal $S_{2\omega}(\tau)$ for the coherent state of Eq. (1) of the main text of the manuscript. The result is shown in Fig. 2a.  However, the interesting observation in the $S_{2\omega}(\tau)$ appears when using the optical GCSS of Eq. (2) of the main text of the manuscript. The significant change in the signal is due to the interference between the two states composing the GCSS leading to beating signatures (see Figs.~2b, c).\\

\noindent\textbf{{Part 8: 2-AC of a coherent IR state for large $\delta\alpha$}}\\
Here we present the second order interferometric autocorrelation (2--AC) trace of a coherent state measured for $|\delta\alpha|=1.44$ when the QS is ``off'' (Fig.S3). The 2-AC trace corresponds to a conventional 2--AC trace of a coherent light pulse. When the QS is ``on'' the trace changes to the trace shown in Fig.2e of the main text.\\

\begin{figure}
    \centering
    \includegraphics[width=0.8 \columnwidth]{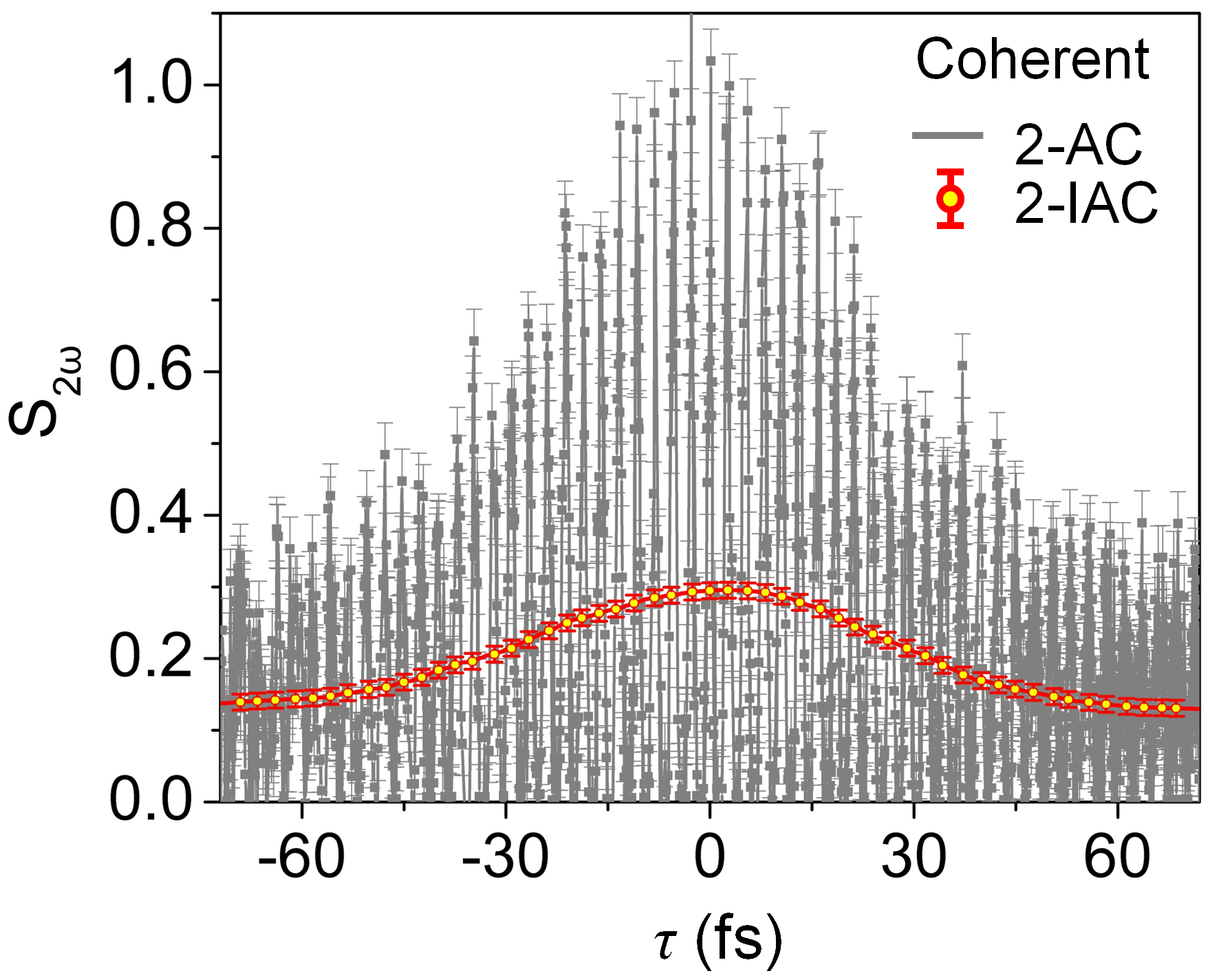}
 \caption{Second order interferometric autocorrelation (2--AC) trace of a coherent state measured for $|\delta\alpha|=1.44$ when the QS is ``off''. The red yellow--filled circles result from the cycle average of the 2-AC traces and they correspond to the second order intensity autocorrelation (2-IAC) traces. The error bar represents one standard deviation of the mean.}
        \label{fig:SM:CoherentHighDa}
	\end{figure}

\noindent\textbf{{Part 9: 2nd order autocorrelation of mixed states}}\\
To verify that the beating structure shown in the 2-AC traces of Figs.~2b, c originates from quantum interference between the two coherent states participating in the superposition for the formation of the GCSS, we have calculated for both values of $\delta\alpha$ shown in Fig.~2, the 2-AC traces and the Wigner functions using the classical mixture of the two coherent states, i.e., for the mixture $\rho(t, \tau) = \dyad{\psi(t,\tau)} + \abs{\bra{\alpha(t)}\ket{\psi(t,\tau)}}^2  \dyad{\alpha(t)}$ (Fig.~S4a--d). The absence of the beating signal in the 2-AC traces of the classical mixture of the two coherent states confirms that the beating features observed in Fig.~2 of the main text of the manuscript arise from the interference term of the GCSS, and thus, has no classical counterpart.\\

\begin{figure}
    \centering
    \includegraphics[width=1 \columnwidth]{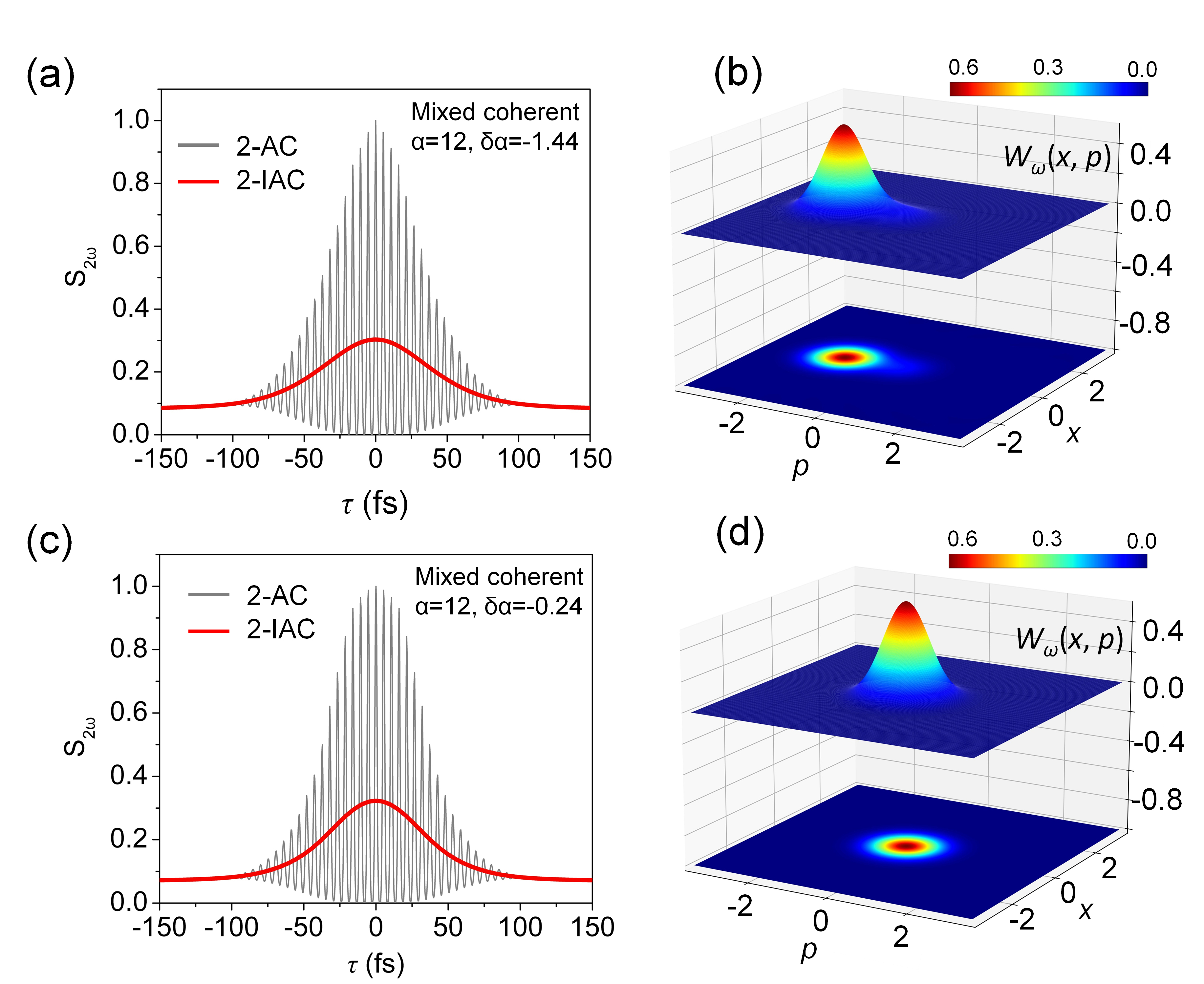}
 \caption{2-AC traces and Wigner functions of mixed IR coherent states. (a), (b) 2-AC trace and the Wigner function $W_{\omega}(x, p)$ in phase space $(x,p)$ of a mixed IR coherent state with $|\delta\alpha|=1.44$. (c), (d) 2-AC trace and the $W_{\omega}(x, p)$ of a mixed coherent state with $|\delta\alpha|=0.24$. The $W_{\omega}(x, p)$ have been centered around the value of $\alpha$. $\it x$, $\it p$ are the values of the non-commuting quadrature field operators $\hat{x}=(\hat{a}+\hat{a}^\dagger)/\sqrt{2}$ and $\hat{p}=(\hat{a}-\hat{a}^\dagger)/i\sqrt{2}$, and $\hat{a}$, $\hat{a}^\dagger$ are the photon annihilation and creation operators, respectively.}
        \label{fig:SM:theory}
	\end{figure}

\noindent\textbf{{Part 10: Generation of optical GCSS with $2\omega$ frequency}}\\
The theoretical results presented in Fig.~4 were obtained by explicitly considering the interaction of the fundamental (infrared) mode, with the BBO crystal. More specifically, the Schrödinger equation describing the dynamics of this interaction is given, in the interaction picture with respect to the free-field Hamiltonian, by
\begin{equation}\label{eq:Schr:SHG}
    i\hbar \pdv{\ket{\Psi(t)}}{t}
        = \Big[
            \chi
            \big(
                \hat{a}_{\omega}^2 \hat{a}^\dagger_{2\omega}
                + (\hat{a}_{\omega}^\dagger)^2 \hat{a}_{2\omega}
            \big)
          \Big]\ket{\Psi(t)},
\end{equation}
where $\hat{a}_{\omega}$ ($\hat{a}^\dagger_{\omega}$) and $\hat{a}_{2\omega}$ ($\hat{a}^\dagger_{2\omega}$) represent the annihilation (creation) operators acting on the fundamental and second harmonic modes, respectively. The $\chi$ parameter denotes the coupling constant of the interaction, which is proportional to the second-order susceptibility of the BBO crystal.

This equation was solved for two different cases, depending on whether the initial input fields exhibited non-classical features or not (see Fig.~2). Specifically, for Fig.~4a, we considered the fundamental to initially correspond to the state of Eq. (1) of the main text of the manuscript, while for Fig.~4b and 4c to be the coherent state superposition presented of Eq. (2) of the main text of the manuscript. In both cases, the initial state of the second harmonic generation modes was set to a vacuum state $\ket{0}_{2\omega}$. More explicitly, the two initial states considered in the dynamical evolution were
\begin{equation}\label{Eq:first:initial}
    \ket{\Psi(t=t_0)}
        = \ket{ \frac{1}{2} (\alpha + \delta \alpha) (f_+ e^{i \omega t_0} + f_- e^{i \omega t_0} )}_{\omega}
        \ket{0}_{2\omega},
\end{equation}
for the first case, while for the second
\begin{equation}\label{Eq:second:initial}
    \ket{\Psi(t=t_0)}
        = \ket{\text{GCSS}(t=t_0,\tau=0)}_{\omega}
           \ket{0}_{2\omega}.
\end{equation}

The differential equation in Eq.~\eqref{eq:Schr:SHG} was numerically solved using the built-in functions of the \texttt{qutip} Python package~\cite{johansson_qutip_2012,johansson_qutip_2013}. For both Eq.~\eqref{Eq:first:initial} and \eqref{Eq:second:initial}, the initial mean photon number was set to be on the order of hundreds of photons ($\lvert \alpha\rvert \approx 12$). Also the time-delay was set $\tau = 0$. We consider this case as it provides the strongest non-classical features for the generated second harmonic. Different values of $\tau \in (0,2\pi/\omega]$ lead to non-optimal interference and reduced non-classical features.

From a more technical side, as the numerical analysis involved expressing states and operators in the Fock basis, which has an infinite number of basis elements, the employed Fock basis needed to be truncated to a certain value $n_{\text{max}}$. To ensure convergence of the results for the considered initial states, it is necessary for $n_{\text{max}} \gg 100$. For the results in Fig.~4, we set $n_{\text{max}} = 500$, which were benchmarked against higher values of this truncation parameter ($n_{\text{max}} = 600$) to ensure convergence of the results. However, higher values of $n_{\text{max}}$ led to matrices that were sufficiently big to exceed the memory capabilities of the employed hardware. 

The value of $\chi$ in Eq.~\eqref{eq:Schr:SHG}, as well as the total interaction time, were adjusted to achieve the mean photon numbers for the generated second harmonic radiation in the range of few to tenths of photons. Propagation effects within the crystal have not been considered as they are out of the scope of this work. In this direction, it is noteworthy that increasing these two quantities results in larger mean-photon numbers for the second harmonic mode, but also in stronger non-classical features when using Eq.~\eqref{Eq:second:initial} as the initial state. This can be considered as an additional knob for controlling the non-classical features of the second harmonic. From a more technical part, to handle the chosen parameters (initial mean photon numbers, susceptibility and total interaction time) numerically within a reasonable timeframe, the use of time-independent Hamiltonians is imperative. Thus, in these calculations, we consider the case of monochromatic field by setting $f(t)=1$. The latter, is as a safe approximation since the generation of the second harmonic mainly occurs around the peak of the pulse where the field amplitude can be considered constant.\\

    \bibliography{References_PRL}{}